\documentclass{nature}

\usepackage[dvipdfmx]{graphicx}
\makeatletter
\let\saved@includegraphics\includegraphics
\AtBeginDocument{\let\includegraphics\saved@includegraphics}
\renewenvironment*{figure}{\@float{figure}}{\end@float}
\makeatother
\usepackage[separate-uncertainty=true]{siunitx}
\usepackage{subcaption}
\usepackage{amssymb}
\usepackage{amsmath}
\usepackage{enumitem}
\usepackage{bbold}
\usepackage{color}
\usepackage{lineno}
\usepackage{ulem}
\usepackage{wasysym}
\usepackage{multirow}
\usepackage{pdflscape}
\usepackage{booktabs}
\usepackage{comment}
\usepackage{physics}
\usepackage[version=3]{mhchem}

\title{Sub-millimeter propagation of antiferromagnetic magnons via magnon-photon coupling}
\author{Ryo~Kainuma$^{1\#}$, Keita~Matsumoto$^{1,2\#}$, Toshimitsu~Ito$^{3}$ and Takuya~Satoh$^{1,4*}$}

\begin{document}
\maketitle
\begin{affiliations}
    \item Department of Physics, Tokyo Institute of Technology, Tokyo, 152-8551, Japan
    \item Department of Physics, Kyushu University, Fukuoka, 819-0395, Japan
    \item National Institute of Advanced Industrial Science and Technology, Tsukuba, 305-8565, Japan
    \item Quantum Research Center for Chirality, Institute for Molecular Science, Okazaki, 444-8585, Japan\\
    $^\#$These authors contributed equally to this study.\\
    $^*$e-mail: satoh@phys.titech.ac.jp
\end{affiliations}

\begin{abstract}
For the realization of magnon-based current-free technologies, referred to as magnonics, all-optical control of magnons is an important technique for both fundamental research and practical applications.
Magnon-polariton is a coupled state of magnon and photon in a magnetic medium, expected to exhibit magnon-like controllability and photon-like high-speed propagation.
While recent studies have observed magnon-polaritons as modulation of incident terahertz waves, the influence of magnon-photon coupling on  magnon propagation properties remains unexplored.
This study aimed to observe the spatiotemporal dynamics of coherent magnon-polaritons through time-resolved imaging measurements.
\ce{BiFeO3} was selected as the sample due to its anticipated strong coupling between magnons and photons.
The observed dynamics suggest that antiferromagnetic magnons can propagate over long distances, up to hundreds of micrometers, through strong coupling with photons.
These results enhance our understanding of the optical control of  magnonic systems, thereby paving the way for terahertz opto-magnonics.
\end{abstract}

\section*{Introduction}
In the past decade, significant research has been conducted on revolutionary information-processing devices that utilize magnons as an information carrier, marking significant progress in the field of magnonics\cite{Kruglyak10,Pirro21}.
Following the study reporting the spin-wave logic gate\cite{Schneider08}, there has been extensive exploration of ferromagnetic magnons.
In recent years, research focus has extended from ferromagnets (FMs) to antiferromagnets (AFMs)\cite{Baltz18,Li20}.
This extention is primarily due to the higher resonance frequencies of AFMs, extending into the terahertz region.
It is desired that magnons are excited coherently and propagate over distances longer than their wavelengths to leverage superposition and nonlinearity\cite{Pirro21, Chumak22}.
However, the excitation and detection of magnons in AFMs pose greater challenges compared to FMs due to the absence of net magnetization in AFMs.

Established methods for magnon excitation include electrical spin injection via the spin Hall effect\cite{Hirsch99} and the use of nanometric coplanar waveguides.
While these methods can be applied to AFMs\cite{Lebrun18,Wang23}, they require microfabrication.
On the other hand, ultrashort light pulses can excite and detect coherent magnons in a noncontact manner without the need for microfabrication.
Previous studies have demonstrated the all-optical excitation and detection of antiferromagnetic magnons, attributing these optical excitation mechanisms to inverse magneto-optical effects\cite{Satoh10,Nemec18}.
Such nonthermal optical excitation is typically limited to addressing magnons near the Brillouin zone center or edge, as magnons in these regions often have a group velocity close to zero in AFMs, hindering their propagation.

The energy transfer between magnons and photons gives rise to their coupled states, known as magnon-polaritons. 
Magnon-polaritons can be optically excited and are expected to exhibit a nonzero group velocity due to their anti-cross-energy dispersion.
Previous studies have primarily focused on the magnonic features of the transmitted terahertz wave, which lacks spatial resolution\cite{Mills74, Grishunin18,Boventer23}.
Consequently, the influence of magnon-photon coupling on magnon propagation properties remains unclear.

This article reports on the all-optical coherent excitation and detection of magnon-polaritons in \ce{BiFeO3}.
This material is anticipated to demonstrate strong magnon-photon coupling due to the presence of electromagnons\cite{Rovillain10}. 
Additionally, the multiferroic nature of \ce{BiFeO3} (ref. \cite{Fiebig16}) enabled the excitation and detection of magnon-polaritons in a single material in contrast to a previous study using a hybrid ferroelectric-antiferromagnetic waveguide\cite{Sivarajah19}.
Time-resolved imaging measurements revealed that antiferromagnetic magnon in \ce{BiFeO3} can propagate over long distances, up to hundreds of micrometers in tens of picoseconds, by forming magnon-polaritons.
The excitation is based on nonthermal mechanism, which does not require optical absorption\cite{Hortensius21}.

\begin{figure}[t]
  \centering
  \includegraphics[width=\linewidth]{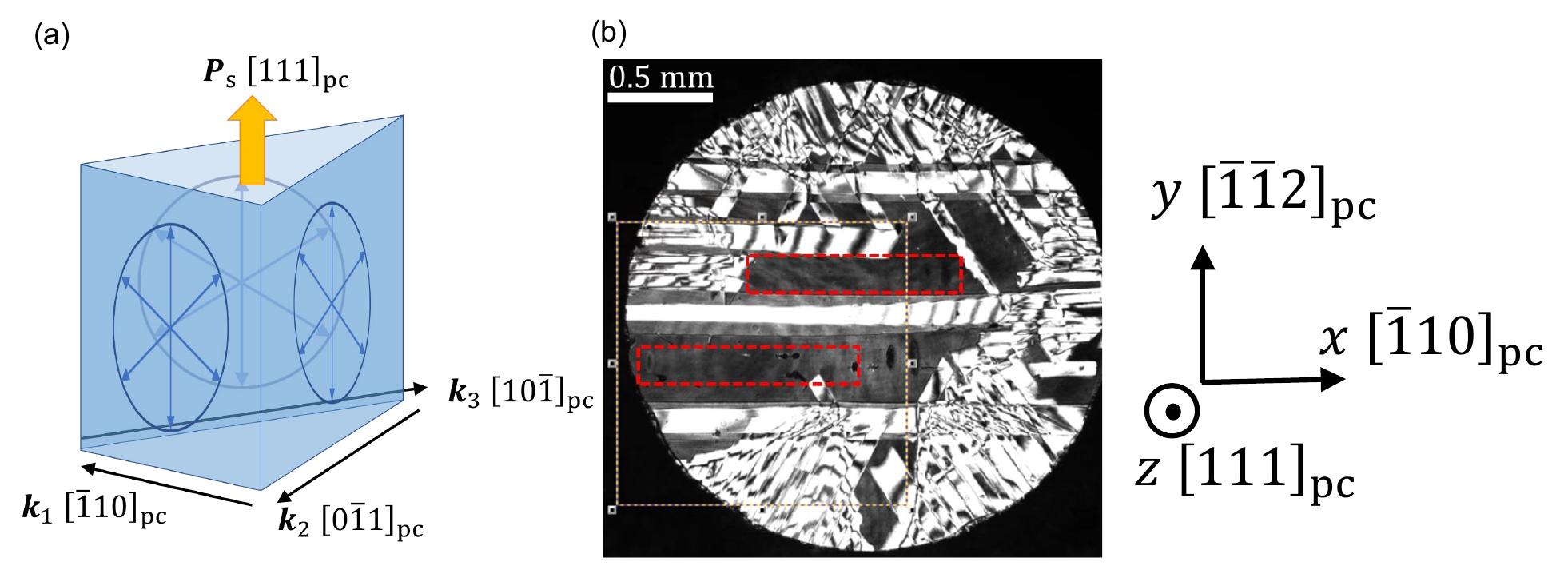}
  \caption{\textbf{Multiferroic \ce{BiFeO3} single crystal.}
  \textbf{a} The schematic structure of multiferroic order of \ce{BiFeO3}.
  The blue bidirectional arrows illustrate the three directional magnetic domains of the cycloidal order.
  \textbf{b} \ce{BiFeO3} single crystal used for our measurements.
  The spatiotemporal waveforms observed in the two red boxes were utilized for the analysis.
  The yellow dashed square represents the area of $1000\times1000$ pixels that the camera measures at a time.}
  \label{fig:Figure_BFO}
\end{figure}

Figure \ref{fig:Figure_BFO}a depicts a schematic of the multiferroic order in \ce{BiFeO3}.
At room temperature, \ce{BiFeO3} exhibits multiferoicity, characterized by ferroelectricity ($T_{\mathrm{C}} = \SI{1100}{K}$) and antiferromagnetism ($T_{\mathrm{N}} = \SI{640}{K}$).
In the antiferromagnetic phase, the N\'eel vector $\mathbf{L}$ displays a cycloidal order with a period of $\SI{62}{nm}$\cite{Sosnowska82, Burns20}.
The wavevector $\mathbf{q}$ of the cycloidal order typically orients in the $[\bar{1}10]_{\mathrm{pc}},\ [0\bar{1}1]_{\mathrm{pc}},\ [10\bar{1}]_{\mathrm{pc}}$ directions, perpendicular to the ferroelectric polarization, $\mathbf{P}_{\mathrm{s}} \parallel [111]_{\mathrm{pc}}$.
Antiferromagnetic magnons, the eigenmodes of $\mathbf{L}$, are classified into $\Phi$ and $\Psi$ modes, corresponding to in-plane and out-of-plane oscillations, respectively, relative to the plane spanned by $\mathbf{P}_{\mathrm{s}}$ and $\mathbf{q}$ (ref. \cite{Sousa08}). %
The spatial periodicity of the cycloidal order folds the magnon Brillouin zone, and deviations can be expanded using modulated plane wave solutions $\exp [-i(\mathbf{k}+n\mathbf{q})\cdot \mathbf{r}]$, where $n$ denotes the $n$th magnon Brillouin zone.
This allows magnons with wavevectors $\mathbf{k}+n\mathbf{q}$ to be accessible via optical measurements, such as Raman\cite{Cazayous08, Rovillain10,Buhot15} or infrared spectroscopy\cite{Komandin10, Talbayev11, Skiadopoulou15, Matsubara16, Bialek18, Farkas21} and pump-probe measurements\cite{Khan20}.
Accordingly, the magnon eigenmodes are labelled as $\Phi_n$ and $\Psi_n$.
Additionally, eigenmodes split by higher-order perturbations are classified using subscripts, as in $\Phi_n^{(m)}$ and $\Psi_n^{(m)}$ (ref. \cite{Fishman15}).

\section*{Results}

The \ce{BiFeO3} single crystal was cut along the $(111)_{\mathrm{pc}}$ plane.
Figure \ref{fig:Figure_BFO}b displays the ferroelectric domain structure of the \ce{BiFeO3} sample, observed using a CMOS camera under crossed-Nicols conditions.
The gray regions within the red rectangular frames did not exhibit optical birefringence, indicating that the polarization ($\mathbf{P}_{\mathrm{s}}$) was homogeneously perpendicular to the surface in this area.
The region inside the frame was selected for the observations.

\begin{figure}[t]
  \centering
  \includegraphics[width=\linewidth]{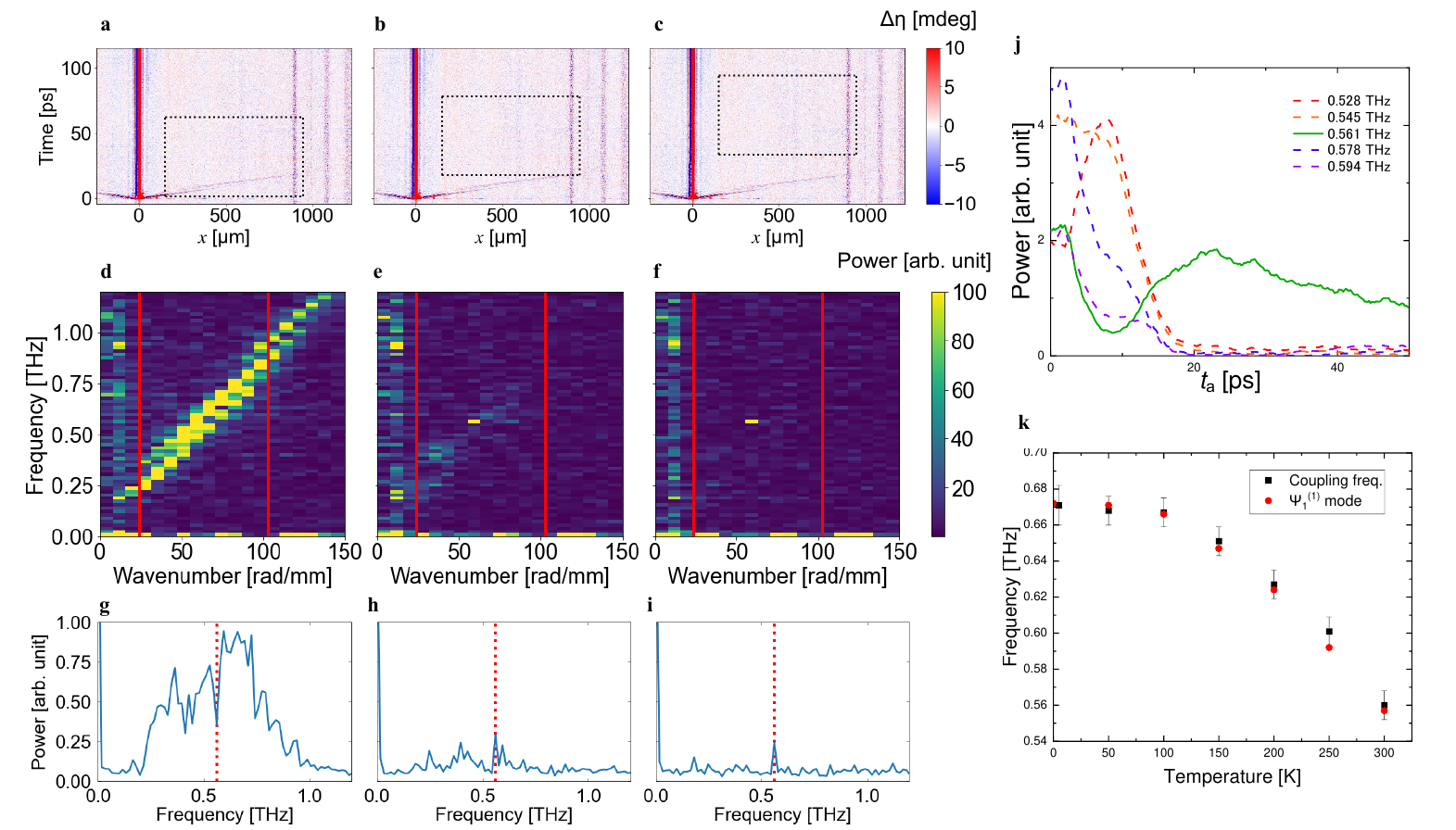}
  \caption{\textbf{Coherent propagation of magnon-polaritons and phonon-polaritons.}
  \textbf{a} The spatiotemporal waveform observed in the pump-probe imaging measurement. 
  The colors in the heat map, ranging from red to blue, represent the ellipticity change of the circularly polarized probe pulse induced at each point of the sample.
  The black dashed frame delineates the region of the Fourier transform for (\textbf{d}).
  \textbf{b}, \textbf{c} shows the same waveform as \textbf{(a)} with different Fourier transform regions for (\textbf{e}, \textbf{f}), respectively.  
  \textbf{g}--\textbf{i} are distributions in (\textbf{d}--\textbf{f}) 
  integrated in the wavenumber direction within the region delimited by the red lines in (\textbf{d}--\textbf{f}).
  The red dashed line in (\textbf{g}--\textbf{i}) represents $\SI{0.56}{THz}$.
  \textbf{j} represents the time evolution of the power of the components around the (anti-)crossing point.
  \textbf{k} illustrates the temperature dependence of the frequency, which exhibited anomalous behavior in (\textbf{j}), along with the eigenfrequency of the magnon $\Psi_1^{(1)}$ mode\cite{Bialek18}. The error bars are defined by the frequency resolutions.}
  \label{fig:Figure_results}
\end{figure}

Figure \ref{fig:Figure_results}a depicts the spatiotemporal waveform  propagating in the $x$-direction, obtained by integrating the entire waveform (see Supplementary Movie SM1) along the $y$-axis.
It is evident that the wavepackets propagated beyond the pump spot at $x=\SI{0}{\micro\meter}$.
These wavepackets corresponded to phonon-dressed terahertz electromagnetic waves, also known as phonon-polaritons\cite{Auston83}.
The excitation mechanism of phonon-polaritions is based on difference-frequency generation\cite{Matsumoto20} (see also Supplementary Information I). 

A Fourier transform was performed to visualize the population of phonon-polaritons and magnon-polaritons in $k$--$f$ space,
employing waveforms ranging from $150$ to $\SI{945}{\micro\meter}$ from the excitation spot (the region within the black dashed frame, as depicted in Fig. \ref{fig:Figure_results}a).
The frequency resolution is primarily limited by the time window (50 ps).
Figure \ref{fig:Figure_results}d illustrates the linear dispersion of phonon-polaritons.
Through this method, components with a finite group velocity were selectively extracted when excluding the excitation spot from the Fourier transform. 
Consequently, Fig. \ref{fig:Figure_results}d  does not contain signals of phonons or magnons localized at the pump spot.
The velocity of the phonon-polaritons was approximately $\SI{6.1e7}{m.s^{-1}}$, leading to a corresponding refractive index of $n_{\mathrm{THz}}=4.9$.
This value aligns with that reported in a previous study [$n_{\mathrm{THz}} = 5.2$ at $\SI{400}{K}$ (ref. \cite{Bialek19})], 
and it surpasses the refractive index for the pump pulse ($n_{\mathrm{pu}} = 2.7$).
Note that the splitting of the phonon-polariton mode is visible in Fig. \ref{fig:Figure_results}d, which may be attributed to the multimodal nature of the phonon-polariton with its finite thickness\cite{Yang10}.
However, anticrossings corresponding to magnon-photon coupling were not clearly identified.

We translated the Fourier transform domain in the time direction to selectively extract the components with different group velocities, as depicted in Fig. \ref{fig:Figure_results}a--c.
Interestingly, we observed that a specific component at $f=\SI{0.56}{THz}$ and $k=\SI{60}{rad.mm^{-1}}$ exhibited a significant increase in power later than the other components, as illustrated in Fig. \ref{fig:Figure_results}d--f in sequence (refer to Supplementary Movie SM2).
Subsequent temperature-dependence measurements revealed that the frequencies at which these specific components appeared were consistent with the frequencies of the magnon $\Psi_1^{(1)}$ mode\cite{Talbayev11,Skiadopoulou15,Matsubara16,Bialek18,Khan20}, as shown in Fig. \ref{fig:Figure_results}k (see also Supplementary Information II for the distribution power spectra).
Hence, we conclude that this signal is located at the anticrossing point of the dispersion curves of magnon $\Psi_1^{(1)}$ mode and photons, indicating that the signal originates from magnon-polaritons.

To extract additional information from the heat maps (Fig. \ref{fig:Figure_results}d--f), 
we first integrated the excitation intensities near the anticrossing in the $k$-direction, resulting in Fig. \ref{fig:Figure_results}g--i.
Figure \ref{fig:Figure_results}g reveals a dip at $\SI{0.56}{THz}$, supporting the notion that the magnon-polariton exhibits a relatively slower group velocity compared to the other nearly non-perturbated photons.
Additionally, Fig. \ref{fig:Figure_results}h, i illustrate that the magnon-polaritons reached the Fourier transform region after the other components had passed through.
For further confirmation, we plotted the time evolution of the powers of several frequency components at $k=\SI{60}{rad.mm^{-1}}$, as shown in Fig. \ref{fig:Figure_results}j.
The horizontal axis represents the start time ($t_a$) of the time domain used in the Fourier transform.
The solid and dashed lines represent the time evolution of the power of each frequency component, corresponding to the change in pixel brightness in Fig. \ref{fig:Figure_results}d--f.
As anticipated, a noticeable difference in behavior was observed, with the magnon-polaritons of $\SI{0.56}{THz}$ component increasing later than those of the other components.
These observations confirmed that magnon-polariton exhibit a group velocity in the intermediate region between the non-perturbated magnons and photons (phonon-polaritons).

Magnon-polaritons near the anticrossing point exhibited group velocities in the range of $0$--$\SI{6.1e7}{m.s^{-1}}\ (=c/n_{\mathrm{THz}})$ depending on the wavenumbers. 
Assuming an average group velocity of $\SI{3.0e7}{m.s^{-1}}$ for the excited magnon-polariton, as a value in the middle of this range, it can be concluded that the group velocity of the magnon-polariton is three orders of magnitude higher than that in the conventional ferromagnetic system \cite{Mahmoud20}.

\section*{Discussion}
In general, the magnon dispersion of antiferromagnets is described as follows \cite{Hortensius21}:
\begin{equation}
\omega=\sqrt{\omega_0^2+(v_0 k)^2}
\end{equation}
For orthoferrite in Ref. \cite{Hortensius21}, $\omega_0 \approx 2\pi \times 175 \times \SI{e9}{rad.s^{-1}}$ and $v_0 \approx 2 \times \SI{e4}{m.s^{-1}}$. The magnon group velocity $\partial \omega/\partial k$ of $1.3 \times \SI{e4}{m.s^{-1}}$ was obtained at a magnon wavenumber of $k=4.2 \times \SI{e4}{rad.mm^{-1}}$. The propagation length can be calculated as the product of the group velocity and the lifetime. With a lifetime of $\SI{85}{ps}$, this gives rise to a propagation length of $\sim \SI{e-6}{m}$.
For \ce{BiFeO3} in Ref. \cite{Cazayous08}, the magnon group velocity of $1.4 \times \SI{e4}{m.s^{-1}}$ was obtained at the magnon wavenumber of $k=q=2\pi/(\SI{62}{nm}) \approx \SI{e5}{rad.mm^{-1}}$. In both cases, the group velocity is $\sim \SI{e4}{m.s^{-1}}$ at a wavenumber of $\sim \SI{e5}{rad.mm^{-1}}$. It is important to note that these velocities correspond to pure magnons, not to coupled magnon-polaritons.
In our case, the excited wavenumber of the magnon-polariton is $k \approx \SI{60}{rad.mm^{-1}}$. Without coupling to phonon-polariton, the group velocity of the $\Psi_1^{(1)}$ magnon at this wavenumber would be $\sim \SI{e4}{m.s^{-1}}$ for $\mathbf{k} \parallel \mathbf{q}$ and $\sim \SI{10}{m.s^{-1}}$ for $\mathbf{k} \perp \mathbf{q}$ \cite{Sousa08}. The lifetime of $\sim \SI{100}{ps}$ (Ref. \cite{Khan20}) leads to the propagation length of $\sim \SI{e-6}{m}$ for $\mathbf{k} \parallel \mathbf{q}$ and $\sim \SI{e-9}{m}$ for $\mathbf{k} \perp \mathbf{q}$.
However, when magnons are coupled to phonon-polaritons with a velocity of $c/n_{\mathrm{THz}}= 6.1\times \SI{e7}{m.s^{-1}}$, as observed in our study, the group velocity of the magnon-polariton is significantly enhanced at the anticrossing point, reaching an average of $3 \times \SI{e7}{m.s^{-1}}$. The lifetime of $\SI{40}{ps}$ (as shown in Fig. \ref{fig:Figure_results}j) gives rise to the propagation length of $\sim \SI{e-3}{m}$.
The sub-millimeter propagation of magnon-polaritons is also evident from Fig. \ref{fig:Figure_results}d--f, obtained by Fourier transform from 150 to $\SI{945}{\micro m}$, particularly with the presence of the spot at $k \approx \SI{60}{rad.mm^{-1}}$ associated with magnon-polaritons in Fig. \ref{fig:Figure_results}e, f.

\section*{Conclusion}
This study conducted time-resolved imaging measurements of multiferroic \ce{BiFeO3} and observed that antiferromagnetic magnon can propagate over hundreds of micrometers via magnon-photon coupling.
The analyses based on Fourier transform provided wavenumber-resolved information that is inaccessible by conventional terahertz spectroscopy.
Additionally, the refractive index of \ce{BiFeO3}, $n_{\mathrm{THz}}$, was extracted, and approximate value of the group velocity of magnon-polaritons was determined.
Although the data were obtained using an unprocessed single crystal, future extensions such as coupling enhancement by a cavity opto-magnonic system are feasible\cite{Parvini20}.
Thus, our findings offer comprehensive insights into antiferromagnetic magnonics, polaritonics, and potential future applications.

\section*{Methods}
The $\SI{120}{\micro\meter}$-thick $(111)_{\mathrm{pc}}$-oriented \ce{BiFeO3} single crystal was grown using the modified floating-zone method with laser diodes\cite{Ito11}, employing magnetic annealing in the $[10\bar{1}]_{\mathrm{pc}}$-direction.
Pump-probe imaging measurements for \ce{BiFeO3} were conducted using femtosecond laser pulses.
A Ti:sapphire regenerative amplifier with a repetition rate of $\SI{1}{kHz}$ and a pulse duration of $\tau = \SI{60}{fs}$ was utilized to generate both the pump and probe pulses with a central wavelength of $\SI{800}{nm}$. 
Additionally, the central wavelength of the pump pulse was converted to $\SI{1300}{nm}$ using an optical parametric amplifier.
Linearly polarized pump pulses with a wavelength of $\SI{1300}{nm}$ were focused onto the sample as a line-shaped spot with a width of $\SI{20}{\micro\meter}$ using a cylindrical lens. 
The pump pulses were directed to the left edge of the region depicted in Fig. \ref{fig:Figure_BFO}b.
The azimuthal angle of the pump polarization was adjusted using a half-wave plate.
Additionally, the probe pulse with a wavelength of $\SI{800}{nm}$ was irradiated without focusing and with varying time delays relative to the pump pulse.
The probe pulse was circularly polarized using a quarter-wave plate (QWP).
Moreover, the probe pulses became elliptically polarized due to the sample's electro-optical (EO) or magneto-optical effect.
The transmitted probe pulse was then transformed into an approximately linearly polarized pulse using another QWP.
Additionally, the ellipticity change of the probe pulse was determined using a rotating-polarizer method with a wire grid analyzer and a CMOS camera\cite{Yoshimine14}.
An optical configuration sensitive to the off-diagonal component of the refractive index modulation was employed to selectively extract extraordinary electromagnetic waves\cite{Matsumoto20}.
The detection mechanism is explained in Supplementary Information III.
The sample was placed in a cryostat for temperature-dependent measurements ranging from $5$ to $\SI{300}{K}$.

\section*{Acknowledgements}
We thank K. T. Yamada for the valuable discussions and technical support.
This study was financially supported by the Japan Society for the Promotion of Science (JSPS) KAKENHI (Grant Nos. JP19H01828, JP19H05618, JP19J21797, JP19K21854, JP21H01032, JP22H01154 and JP22J20939), the Frontier Photonic Sciences Project (NINS Grant Nos. 01212002 and 01213004), OML Project (NINS Grant No. OML012301) of the National Institutes of Natural Sciences (NINS) and MEXT Initiative to Establish NeXt-generation Novel Integrated Circuits CenterS (X-NICS) (Grant No. JPJ011438).

\section*{Author contributions}
T.S. conceived and supervised the study. R.K., K.M., and T.S. performed the experiments and analyzed the data. T.I. fabricated samples. R.K. and T.S. wrote the manuscript. All authors discussed the results and provided comments on the manuscript.

\section*{Data availability}
The datasets generated during and/or analyzed during the current study are available from the corresponding author upon reasonable request.

\section*{Competing interests}
The authors declare no competing interests.

\section*{References}
\bibliographystyle{naturemag}

\newpage
\newpage

\newpage

\end{document}